\documentstyle[aps,prl]{revtex}

\begin{document}

\input epsf.sty

\draft

\twocolumn[\hsize\textwidth\columnwidth\hsize\csname %
@twocolumnfalse\endcsname

\title{NMR Spin-Spin Relaxation as Kinetics in Spin Phase Space}

\author{Boris V. Fine}
\address{Department of Physics, University of Illinois at
Urbana-Champaign, Urbana, IL 61801}

\date{July 17, 1997}

\maketitle

\begin{abstract}
A new approach is presented that treats 
NMR spin-spin relaxation as kinetics in spin phase space. 
The approach is applied to 
free induction decay (FID) in solids containing equivalent nuclear spins 1/2.
The description obtained does not
involve adjustable parameters.
As an example, the calculation is performed for $^{19}$F FID in CaF$_2$, 
and the results are in
good agreement with experiment.
\end{abstract}
\phantom{.}

]

\narrowtext
\pagebreak

 We present a new approach to the problem of
free induction decay (FID)
in the lattice of  equivalent spin $1/2$ nuclei,
where nuclear spin-spin interaction is responsible for the relaxation.
The approach applies the framework of the Boltzmann
kinetic description beyond the limit of instantaneous collisions.
The resulting method is formally
comparable to the original 
treatment of Lowe and
Norberg~\cite{LN}, in the sense that it starts from the Ising-like
Hamiltonian, matches the second and the fourth momenta, 
and does not involve adjustments to
experimental data. Moreover, the T-criterion, which is introduced
in this work, is also satisfied in the Lowe and Norberg calculation. 
The difference is that our mathematical construction is  
based on physical arguments, which give  better control
over the calculation, and, in principle, allow
a routine analysis of experiments to be performed  to extract unknown
microscopic parameters. None of the approximate methods suggested
so far has been generally accepted for such a task. 

A typical FID formulation in solids assumes that each nucleus is at rest
on its site in the crystal lattice, 
and  the following conditions are fulfilled:
\begin{equation}
{k_B T \over \hbar} >\! > \Omega >\! > {1 \over T_2} >\! > {1 \over T_1},
\label{assumptions}
\end{equation}
where
$T$ is the temperature
of the initial equilibrium distribution,
$\Omega$ is the Larmor frequency in the static magnetic field,
$T_2$ is the 
time scale of the nuclear spin-spin
interaction, 
and $T_1$ is the spin-lattice relaxation time.  
The spin-lattice relaxation can be neglected,
since the time range of interest is of the order of $T_2$. 

We consider FID in the Larmor rotating reference frame, where the
Hamiltonian  is \cite{VV}
\begin{equation}
{\cal H} = \sum_{k<n}  
[A_{kn} I_{kz} I_{nz} + B_{kn} \hbox{{\boldmath $I$}}_k \cdot 
\hbox{{\boldmath $I$}}_n], 
\label{Htr}
\end{equation}
$\hbox{{\boldmath $I$}}_k$ is the spin 1/2 operator of the 
$k$th nucleus, and $A_{kn}$ and $B_{kn}$ are the interaction coefficients.
The $z$-axis is chosen along the direction of the Larmor precession.

The FID is measured as a spin response to 
$ \pi/2 $ radio frequency pulse. The $ \pi/2 $ pulse rotates 
the equilibrium spin
system so that immediately after the
pulse, all spins are uniformly polarized in the direction
perpendicular to the static magnetic field. We choose
the  $ x $-axis  along 
the magnetization at this moment of time.
After the pulse,  
the $x$-component of the magnetization $M_x$ decays 
as a result of the spin-spin interactions. 
The leading term in the high temperature expansion gives 
\cite{LN,Abr}:
\begin{equation}
M_x (t) = {\gamma \hbar^2 \Omega \over k_B T} \ \ 
\hbox{Tr}\{ e^{ i \> {\cal H}  \>  t } 
                                     \> \sum_{k} I_{kx}
                                     \; e^{ - i \>  {\cal H}  \> t }
                                     \> \sum_{n} I_{nx}
                              \},
\label{Mperp}
\end{equation}
where $\gamma$ is the gyromagnetic ratio.
Usually, FID is presented as a normalized 
function $G(t) = M_x (t) / M_x (0)$.

Since all spins are equivalent, each of them
equally contributes to the magnetization. 
For the purposes of the forthcoming consideration,
we  express $G(t)$ in
terms  of the average magnetic moment {\boldmath $m$}
of {\em one} spin:  
\begin{equation}
G(t) = m_x(t) / m_x(0).
\label{F}
\end{equation}

We outline our approach
by comparing it with the derivation of the Boltzmann
kinetic equation (BKE) \cite{LP}. 

The BKE can be considered as a modification of the exact equation for the
system of noninteracting classical particles.  
In our method, we modify the exactly soluble case
of the  Ising-like Hamiltonian.
The BKE is derived
 by counting particles
entering and leaving small volumes  of the phase space.
Similarly, we analyze the averaged spin behaviour 
in a given configuration of neighbors.
The configuration of neighbors refers to the phase space domain arising
in the context of 
the Ising-like Hamiltonian. 

The derivation of BKE is based on the 
approximation of intantaneous collisions,
which is not applicable to 
the nuclear spin-spin interaction in solids.
Instead, we  substantiate the
quantitative claim of our approach by adopting the
criteria presented  below. 

The first criterion is that the mathematical construction of
our theory has to be time reversible. This allows
a well-defined correspondence to be established between the parameters of
the theory and the time reversible microscopic dynamics. 
In particular, it enables us to match the exactly calculated 
momenta $M_2= -{d^{2} G \over dt^2} |_{_{t\, =0}}$ and 
$M_4 = {d^{4} G \over dt^4} |_{_{t\, =0}}$, which can be obtained by
the direct trace evaluation of the first terms in
the time expansion of Eq.(\ref{Mperp}).
Time reversibility implies abandoning simple first-order differential rate
equations. Consequently, the minimal description has to be based on 
time-reversible second-order differential rate equations. 

 Among the various dynamical correlations that complicate the analysis,
the most important  are the two-spin
correlations.  If an approximate calculation can accurately take
the two-spin correlations into account, then it is reasonable to expect
that the contribution from the higher order correlations is more
random, i.e better averageable. The criterion,
which at least partially guarantees that the two-spin correlations
are respected, is as follows:
The  FID shape obtained by the effective calculation 
is the function of the microscopic
coefficients $A_{kn}$ and $B_{kn}$. We require
that if in this function all $A_{kn}$ and $B_{kn}$, except
for the coefficients describing the interaction between any given 
pair of spins, equal 
zero, then the function reproduces the exact FID shape of the two-spin system
\begin{equation}
G(t)=cos \left(\hbox{${1 \over 2 \hbar}$} A_{12} t \right) .
\label{GT}
\end{equation} 
We call this the ``T-criterion."

The FID can be evaluated in a closed
form in the  Ising-like case when all $B_{kn} = 0$.
We use the Lowe and Norberg  interpretation \cite{LN} of this
evaluation to initially motivate  the formalism of our
description. 

In the Ising-like Hamiltonian, the operator of the local field,
which affects the $k$th spin, is
\begin{equation}
h_k = \hbox{${1 \over \gamma \hbar}$} {\sum_{n}} A_{kn} I_{nz} \ .
\label{local}
\end{equation}
Since the z-component of each spin is the constant of motion,
Eq.(\ref{local}) implies that each spin is rotated by a constant local
field created by spin's neighbors.

If all operators $I_{nz}$ are diagonal in the basis chosen for the
trace evaluation in Eq.(\ref{Mperp}), the contributions to the trace
can be interpreted as a result of spin precessions in classical local fields.
The possible values of these fields are the eigenvalues of the field
operator $h_k$.

We introduce index ${\cal C}$ to refer to the configurations of 
neighbors, and define configuration as  a particular set of the
eigenvalues of operators $I_{nz}$ in Eq.(\ref{local}).  The description
can be restricted  to a  finite number of neighbors $N_0$.  
Consequently, there are
infinitely many spins surrounded by a given configuration ${\cal
C}$ of $N_0$ neighbors. In the following we use
notation $\langle ... \rangle_{_{\cal C}}$ to indicate
averaging over all configurations.

For configuration  ${\cal C}$, we  define $m_{{\cal C}x}(t)$  as the 
average magnetic moment of nuclei that are surrounded by neighbors
with the specified set of instantaneous spin projections on the $z$-axis. 
The initial uniform polarization implies that
 $m_{{\cal C}x}(0) = m_x(0) = {\gamma \hbar^2 \Omega
\over 4 k_B T}$.

 The evolution of $m_{{\cal C}x}$ is governed by the equation
\begin{equation}
{d^2 m_{{\cal C}x} \over d t^2} = - \omega_{\cal C}^2  m_{{\cal C}x} ,
\label{precession}
\end{equation} 
where ${\omega}_{\cal C}$ is 
the precession rate in configuration ${\cal C}$:
\begin{equation}
{\omega}_{\cal C} = \hbox{${1 \over 2 \hbar}$} \sum_{n} \pm A_{kn}.
\label{omegaC}
\end{equation} 
Each configuration
of neighbors uniquely specifies the combination of signs in Eq.(\ref{omegaC}).

As a result, 
$m_{{\cal C}x}(t) = m_{{\cal C}x}(0) cos({\omega}_{\cal C}t)$,
and Eq.(\ref{F}) with $m_x(t) = \langle m_{{\cal C}x}(t) \rangle_{_{\cal C}}$ 
leads to \cite{LN}
\begin{equation}
G(t) = {\prod_n} cos \left( \hbox{${1 \over 2 \hbar}$} A_{kn} t \right).
\label{product}
\end{equation} 

The right-hand side (RHS) of Eq.(\ref{product}) is 
the Fourier transform of the discrete 
distribution
of ${\omega}_{\cal C}$, which  can be considered
as a sum of random quantities $\pm  A_{kn}/(2 \hbar)$. Therefore,
according to the central limit theorem, the distribution approaches Gaussian
when
\begin{equation}
max\{ |A_{kn}| \} <\! < 2 \hbar \sqrt{M_2},
\label{criterium}
\end{equation}
where $M_2 = {1 \over 4 {\hbar}^2} {\sum_{n}}  \> A_{kn}^2 $. 
In this case, the RHS of Eq.(\ref{product}) can be approximated as
$ exp(-  M_2 \> t^2/2)$.

Apart from the Ising-like case,
the evolutions of $m_{{\cal C}x}$ in different 
configurations of neighbors 
are mutually dependent.
We employ the following generalization of Eq.(\ref{precession}) 
\begin{equation}
{d^2 m_{{\cal C}x} \over d t^2} = - (W_{{\cal C}+}^2  P_{{\cal C}x+} 
- W_{{\cal C}-}^2  P_{{\cal C}x-}) m_0,
\label{rate_eq}
\end{equation} 
where $P_{{\cal C}x+}$ and $P_{{\cal C}x-}$ are the probabilities that
the spin projections on the $x$-axis are positive and negative,
respectively; 
$W_{{\cal C}+}$ is the effective rate of the transition 
from a state where the spin is oriented positively  along
the $x$-axis  to the state where the spin is oriented negatively along the
same axis; $W_{{\cal C}-}$ is the rate of the reverse transition;
$ m_0 = \gamma \hbar /2$ is 
the maximum magnetic moment of one spin.
 The probabilities 
  obey the relationships: 
\begin{eqnarray}
P_{{\cal C}x+} + P_{{\cal C}x-} &=& 1, \label{Psum} \\
m_0 ( P_{{\cal C}x+} - P_{{\cal C}x-} ) &=& m_{{\cal C}x}.
\label{probab}
\end{eqnarray}

The rates in Eq.(\ref{rate_eq}) characterize the net effect of two
factors: the direct influence of neighbors and the flux of spin
polarization to and from configuration ${\cal C}$.
When the spin system is weakly polarized, each of these factors 
can lead to a slight asymmetry between
$W_{{\cal C}+}$ and $W_{{\cal C}-}$.

The first factor can be  
understood after we rewrite the two-spin Hamiltonian extracted from 
Eq.(\ref{Htr}) as
\begin{eqnarray}
{\cal H}_{kn} &=& B_{kn} I_{kx} I_{nx}  
- {1 \over 4} A_{kn} ( I_{k+} I_{n+} +  I_{k-} I_{n-}) \nonumber \\
&& +  \; {1 \over 4} (A_{kn} + 2 B_{kn}) ( I_{k+} I_{n-} +  I_{k-} I_{n+}),
\label{H2}
\end{eqnarray}
where $ I_+ = I_y + i I_z $, and $ I_- = I_y - i I_z $.
The second and the third terms in Eq.(\ref{H2}) lead to
the double-flip of two parallel spins and the flip-flop of two
antiparallel spins, respectively.  
If the $k$th spin is parallel to the
average spin polarization, it is more probable that  the
$n$th spin will be parallel to the $k$th spin. 
This effectively increases the influence 
of the double-flip
term on the $k$th spin and reduces the influence of the flip-flop term.  
If the $k$th
spin is antiparallel to the average polarization, the opposite effect
would occur.  
When $m_x > 0$, the
double-flip correlation increases $W_{{\cal C}+}^2$ and
reduces  $W_{{\cal C}-}^2$
in Eq.(\ref{rate_eq}). The
flip-flop correlation leads to the opposite result.

The other factor mentioned above is frequently referred to as
motional narrowing. In terms of our description, this means that
the faster the time variations of the real local fields, 
the greater the difference between the chosen configuration of neighbors
and those configurations that actually drove the spin to its current 
surroundings.
One can conclude that spins coming to configuration ${\cal C}$ 
from other configurations tend
to be slightly polarized in the direction of the average
magnetization.  This reduces $W_{{\cal C}+}^2$ and increases
$W_{{\cal C}-}^2$, provided $m_x > 0$.

We define the average rate as 
\begin{equation}
W_{{\cal C}}^2 = {1 \over 2} (W_{{\cal C}+}^2 + W_{{\cal C}-}^2),
\label{av_rate}
\end{equation}
The high temperature condition guarantees that the difference between 
$W_{{\cal C}+}^2$ and $W_{{\cal C}-}^2$ is small.
The basic assumption of our analysis is
that the leading term of this difference can be expressed as
\begin{equation}
{1 \over 2} (W_{{\cal C}+}^2 - W_{{\cal C}-}^2) =
     \alpha \; W_{{\cal C}}^2 \; {m_x \over m_0},
\label{assumption}
\end{equation}
where $\alpha$ is a configuration independent parameter, which is
determined by the average of all factors considered. 
The status of assumption (\ref{assumption}) is somewhat similar to the
self-consistent relaxation time approximation, which is
frequently adopted to solve BKE.

The substitution of 
Eqs.(\ref{Psum},\ref{probab},\ref{av_rate},\ref{assumption})
into Eq.(\ref{rate_eq}) yields
\begin{equation}
{d^2 m_{{\cal C}x} \over d t^2} = - W_{\cal C}^2 \;  m_{{\cal C}x} 
- \alpha \; W_{\cal C}^2 \; m_x .
\label{basic}
\end{equation}

Given the initial uniformly polarized state, the solution of Eq.(\ref{basic}) is
\begin{eqnarray}
m_{{\cal C}x}(t) &=& m_{{\cal C}x}(0) \; cos(W_{\cal C} t)  \nonumber \\ 
&&- \  \alpha \ 
{\int_0}^t \! m_x(t-t^{\prime}) W_{\cal C} 
sin(W_{\cal C} t^{\prime}) d t^{\prime}.
\label{mCxt}
\end{eqnarray}

Since $m_x(t) = \langle m_{{\cal C}x}(t) \rangle_{_{\cal C}} $,
the averaging of Eq.(\ref{mCxt}) over all configurations, together with
Eq.(\ref{F}),
results in the integral equation
\begin{equation}
G(t) = g(t) + \alpha {\int_0}^t G(t-t^{\prime}) 
{d g(t^{\prime}) \over d t^{\prime}} d t^{\prime},
\label{result}
\end{equation}
where 
$g(t) = \langle cos(W_{\cal C} t)\rangle_{_{\cal C}}$.

We choose $g(t)$ and $\alpha$ 
such that
the second and the fourth momenta obtained from  Eq.(\ref{result}) 
match  the exact calculation.
Matching the second moment gives
\begin{equation}
M_{2g}  = {M_2 \over 1 + \alpha}, 
\label{m2}
\end{equation}
where $M_{2g}= -{d^{2}\! g \over dt^2} |_{_{t\, =0}} = 
\langle W_{\cal C}^2 \rangle_{_{\cal C}} $.  Matching
the fourth moment, combined with Eq.(\ref{m2}), allows $\alpha$ to be expressed 
as 
\begin{equation}
\alpha = { {M_{4g} \over M_{2g}^2} - {M_{4} \over M_{2}^2} \over 
{M_{4} \over M_{2}^2} - 1},
\label{alpha}
\end{equation}
where $M_{4g}= {d^{4}\! g \over dt^4} |_{_{t\, =0}} = 
\langle W_{\cal C}^4 \rangle_{_{\cal C}} $.

 Eqs.(\ref{m2},\ref{alpha}) relate the shape and the scale
of the acceptable distributions of $W_{\cal C}$. 
Namely, the  shape specifies
the ratio of $M_{4g} / M_{2g}^2$, which enters Eq.(\ref{alpha}) and
determines the value of $\alpha$. With a known $\alpha$,  Eq.(\ref{m2})
gives the value of $M_{2g}$, which defines the scale of the
distribution.

An important property of Eqs.(\ref{m2},\ref{alpha}) is that they
guarantee the fulfillment of the \mbox{T-criterion} 
for any  shape of 
the distribution of $W_{\cal C}$ with finite moments.
The exact solution (\ref{GT}) for the system of two spins $1/2$   
always gives the ratio 
$M_{4} / M_{2}^2 = 1$, which formally leads to an infinite value of
$\alpha$ in Eq.(\ref{alpha}). 
The divergence does not appear in the solution of Eq.(\ref{result}),
because, according to Eq.(\ref{m2}), it is offset by the 
small value of $M_{2g}$. When the two-spin problem is considered as a
limit of the many-spin problem with $A_{kn}, B_{kn} \to 0$, except for 
$A_{12}$ and $B_{12}$,
the solution of Eq.(\ref{result}) converges to the RHS of Eq.(\ref{GT}), 
independently of
the shape of the distribution of $W_{\cal C}$.

The relevance of the T-criterion is supported by the fact that the  
observable part of $G(t)$ obtained from 
Eqs.(\ref{result},\ref{m2},\ref{alpha}) is weakly
sensitive to the variations of the input shape of  
the $W_{\cal C}$ distribution.
Thus relatively crude assumptions about this
shape still should lead to good accuracy in the result. 

Since the fulfillment of the T-criterion is guaranteed, and 
the
important many-spin correlations are
taken into account by the choice of  parameter $\alpha$,
we assume that the distribution of $W_{\cal C}$ is produced 
by uncorrelated contributions from spin neighbors.
As a result, this distribution has a tendency  to be Gaussian.
However, in the presence of a few strongly interacting neighbors, it
can have a peaked structure originating from the Ising-like part of
Hamiltonian (\ref{Htr}). 
This part alone would produce
the discrete distribution of the static rates 
$\omega_{\cal C}$ given by Eq.(\ref{omegaC}). 
The discrete structure, presumably, propagates to the distribution
of $W_{\cal C}$, but it has to be ``washed out" by the polarization
flux between different configurations.
The proper scale this ``lifetime'' effect is given by 
\begin{equation}
M_{2f} = -  \left. {1 \over f(0)} \; {d^2 f(t) \over d t^2} \right|_{t=0},
\label{m2f}
\end{equation}
where $f(t) = 
\hbox{Tr}\{    e^{ i \> {{\cal H}}  \>  t } 
            \> h_k
            \; e^{ - i \>  {{\cal H}}  \> t }
            \> h_k
         \}$.

We choose the input shape of the $W_{\cal C}$ distribution  to be
the convolution of the distribution of $\omega_{\cal C}$ 
with a Gaussian that has the second moment $M_{2f}$. 
Based on this shape,
$g(t)$ becomes a renormalized product of a Gaussian 
and the RHS of Eq.(\ref{product}):
\begin{equation}
g(t) = exp[ - \hbox{${1\over 2}$} M_{2f}  (\eta t)^2 ] \ 
{\prod_n} cos \left( \hbox{${1 \over 2 \hbar}$} A_{kn} \eta  t \right),
\label{g}
\end{equation} 
where $\eta = \sqrt{{M_2 \over (1+ \alpha) (M_2 + M_{2f})}}$ is the
renormalization factor required by Eq.(\ref{m2}).
If the large number of neighbors criterion (\ref{criterium}) is satisfied,
g(t) becomes Gaussian. 

Eqs. (\ref{result},\ref{alpha},\ref{m2f},\ref{g}) form a closed
system, which  allows $G(t)$ to be calculated.

We apply the method to CaF$_2$, where FID is
observed on $^{19}$F nuclei (\mbox{ $\gamma = 25 166.2$ rad
s$^{-1}$ Oe$^{-1}$}), which form a simple cubic lattice having
near-neighbor separation of $d = 2.72325$\AA.  The spin-spin interaction
is assumed to be magnetic dipolar \cite{EL}.

The calculation is performed with the static magnetic
field oriented  along the [100], [110] and [111] crystal directions.
Evaluating $g(t)$ according to Eq.(\ref{g}), we individually include 50
cosines with the interaction constants making the largest contribution
to $M_{2g}$, and approximate the product of other cosines by the
Gaussian exponent, which adds the remaining contribution to $M_{2g}$.
We also approximate $M_{2f}$ by the second moment of one spin correlation 
function 
$\hbox{Tr}\{    e^{ i \> {\cal H}  \>  t } 
            \> I_{nz}
            \; e^{ - i \>  {\cal H}  \> t }
            \> I_{nz}
         \}$,
which is sufficient given the alternating signs
of $A_{kn}$. This gives $M_{2f} = {2 \over 9} M_2$. Variations of 
$M_{2f}$, even by a factor of 2, result in a negligible difference
in the observable part of FID.
In Eq.(\ref{alpha}), we take the exact ratios  $M_4/M_2^2$  
from Ref.\cite{CJ}, and 
 obtain the values  of $\alpha$: 
(0.54, 0.42, 0.42) for the [100], [110] and [111] directions, respectively.

\begin{figure}[t]

\setlength{\unitlength}{0.001cm}
\begin{picture}(6000, 13000)
{
\epsfxsize=4.15in\epsfbox{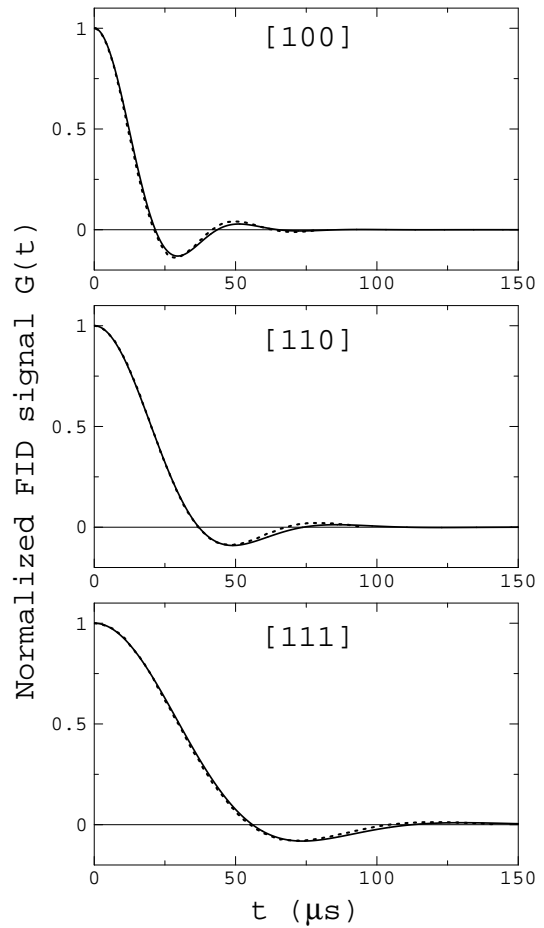}
}

\end{picture}

\caption{ FID
in CaF$_2$ with static magnetic field  along [100], [110] 
and [111] crystal directions. 
Solid lines are the solutions of Eq.(\protect\ref{result}).
Dashed lines are experimental data from Ref.~\protect\cite{EL}.} 
\label{3plots}
\end{figure}

When $g(t)$ and  $\alpha$ are known, Eq.(\ref{result}) is easily
soluble numerically.  The solutions are
plotted in Fig.~\ref{3plots}, together with the 
experimental data of Engelsberg and Lowe \cite{EL}.

The microscopic coefficients originating from the magnetic dipolar interaction 
have the relation 
\mbox{$B_{kn}= - {1 \over 3} A_{kn}$}.  
As a result, the double-flip correlation, which is
discussed after Eq.(\ref{H2}), is sufficiently strong to
outweigh the  flip-flop correlation
and the motional narrowing.
Consequently, $\alpha > 0$, which leads to the
oscillating FID shape.

The value of $\alpha$ can also be negative, e.g.  when 
$A_{kn}$ and $B_{kn}$ have the same sign for each pair of spins. 
If $g(t)$ is Gaussian, and $\alpha$ changes from
$0$ to $-1$, the shape of $G(t)$ obtained from Eq.(\ref{result}) 
evolves from Gaussian to nearly exponential. 
The minimum possible value of $\alpha$ is $-1$, 
which leads to $G(t) = 1$ for any $g(t)$. This limit arises as the 
interaction approaches the Heisenberg form with all $A_{kn}=0$.

In summary, we presented a kinetic FID theory, which reproduced
the exact results of the two-spin problem with Hamiltonian (\ref{Htr})
and  many-spin problem with Ising-like and Heisenberg Hamiltonians. 
Based on the physical arguments,
the theory extended
the above exactly soluble cases 
in the space of all possible interaction
coefficients $A_{kn}$ and $B_{kn}$.  Given also the fact that the first 
\linebreak 
four derivatives  of $G(t)$   at   $t=0$ \ obtained from \  Eq.(\ref{result})

\noindent
coincide with the exact calculation, we conclude that in 
most cases, the solution 
of  Eq.(\ref{result}) approximates the observable part of FID
with  good accuracy. The accuracy is mainly limited 
by the adequacy of assumption (\ref{assumption}). 

The author is grateful to  
C.P.~Slichter and A.~Sokol for stimulating discussions.
This work has been partially supported under Sloan Foundation grant  Br-3438.

\end{document}